\documentclass[prd,twocolumn,showpacs,byrevtex]{revtex4}

\usepackage{graphics}

\newcommand\lsim{\mathrel{\rlap{\lower4pt\hbox{\hskip1pt$\sim$}}
    \raise1pt\hbox{$<$}}}
\newcommand\gsim{\mathrel{\rlap{\lower4pt\hbox{\hskip1pt$\sim$}}
    \raise1pt\hbox{$>$}}}

\def\dddot#1{\mathinner{\buildrel\vbox{\kern5pt\hbox{...}}\over{#1}}}

\def\be{\begin{equation}}
\def\ee{\end{equation}}
\def\bq{\begin{eqnarray}}
\def\eq{\end{eqnarray}}
\def\beq{\begin{eqnarray*}}
\def\eeq{\end{eqnarray*}}

\def\bs{\begin{subequations}}
\def\es{\end{subequations}}
\def\ben{\begin{eqalignno}}
\def\een{\end{eqalignno}}

\def\({\left(}
\def\){\right)}

\begin{document}

\title{Large Angle CMB Fluctuations from Cosmic Strings with a Comological Constant}

\author{M.~Landriau}
\email{landriau@lal.in2p3.fr}
\affiliation{Laboratoire de l'Acc\'{e}l\'{e}rateur
Lin\'{e}aire,\\ IN2P3-CNRS et Universit\'{e} Paris-Sud,\\ B.P. 34, 91898 Orsay
Cedex, France}

\author{E.P.S.~Shellard}
\email{E.P.S.Shellard@damtp.cam.ac.uk}
\affiliation{Department of Applied Mathematics and Theoretical Physics,\\
Centre for Mathematical Sciences, University of Cambridge\\
Wilberforce Road, Cambridge CB3 0WA, U.K.}

\begin{abstract}
In this paper, we present results for large-angle CMB anisotropies
generated from high resolution simulations of cosmic string networks in a
range of flat FRW universes with a cosmological constant.  Using an ensemble
of all-sky
maps, we compare with the COBE data to infer a normalization (or upper
bound) on the string linear energy density $\mu$.  For a flat matter-dominated
model ($\Omega_{M}=1$) we find $G\mu/c^2 \approx 0.7\times 10^{-6}$, which is
lower than previous constraints probably because of the more accurate inclusion 
of string small-scale
structure.  For a cosmological constant within an observationally acceptable
range, we find a relatively weak dependence with $G\mu/c^2$ less than 10\% higher.
\end{abstract}

\pacs{98.80.-k, 98.80.Cq}

\date{August 1, 2003}

\maketitle

\section{Introduction}

Given the growing evidence for a cosmological constant or dark energy
component in the Universe, it is important to determine whether this
significantly impacts the evolution of cosmic strings and their
observational signatures.  The strongest constraints limiting the
energy scale of cosmic strings is currently the COBE normalisation of
large-angle CMB anisotropy.  In this paper, we will infer this
normalisation from maps computed using the methods presented
in~\cite{methods} and determine how this is influenced by
late-time domination by a cosmological constant.

\section{Method}

\subsection{Geometry of Simulations}

To compute all-sky maps, the ``observers'' are located inside the
simulation box.  Following~\cite{pen94},
we place eight such observers at each corner of a cube of side $L/2$
where $L$ is the size of the box itself.  We use the fact that the string
networks, and hence the cosmological perturbations they induce, have
periodic boundary conditions to identify opposite sides of the box.
The eight realizations of the sky are not completely independent, but
the correlations are not apparent.  This scheme reduces the effect of
the cosmic variance on large angular scales and thus renders the
computation of the normalization more accurate.

\subsection{Pixelization}

To facilitate the computation of maps and the extraction of the power
spectrum, we use an iso-latitude pixelization, which enables us to
Fourier transform each ring of constant $\theta$ to evaluate the
integral over the azimuthal angle.  More specifically, we use the
equal area with 3:6:3 base pixels, proposed by~\cite{crittenden98}.
In this scheme, each base pixel is subdivided into $n^2 = 2^{2p}$ pixels.
The area of a pixel is given by $A_{pixel} = 4\pi/N_{pixel} = 
\pi/3n^2$.  But, by integrating the spherical surface element over a
pixel, we have
$A_{pixel} = \frac{2\pi}{N_p}\Delta\cos\theta$.
Hence, the width of a row of pixels is given by
\begin{equation}
\Delta\cos\theta = \frac{N_p}{6n^2} \, ,
\end{equation}
where $N_p$ is the number of pixels per row of constant $\theta$.
In the polar caps, $N_p=3$, then $9\times
2^{i-2}$ repeated $2^{i-2}$ times with $2\leq i \leq p+2$.  The
equatorial regions are made up of $n$ rows of $6n$ pixels.

\subsection{Angular Power Spectrum Computations}\label{cl_comp}

The pixelized map is the convolution of the real temperature map with
the pixel window function.  To compute the power spectrum, we
need to know the spherical transform of the window function.
In real space, the pixel window function is simply given by:
\begin{equation}
W^{pq}(\phi,\theta)  = \left\{\begin{array}{ll}
\frac{1}{A_{pixel}} 	&	\mbox{inside the pixel} \\
0 		&	\mbox{outside} 
\end{array}\right. \, ,\end{equation}
where $p$ and $q$
label the rows and the pixels on that row:
\begin{equation}\label{almcomp}\begin{array}{ll}
T^{pq}
 & = \displaystyle\int_{\mathcal{S}_2}T(\phi,\theta)
 W^{pq}(\phi,\theta)d\Omega \\
 & = \displaystyle\sum_{m=-\ell_{max}}^{\ell_{max}}
 b_me^{\dot{\imath}mq2\pi/N_p} \,,
\end{array}\end{equation}
where
\begin{equation}\label{bmdef}
b_m = \displaystyle\sum_{\ell =|m|}^{\ell_{max}}
a_{\ell m} w(\pi m/N_p)\overline{\lambda_{\ell}^m}  \, ,
\end{equation}
with
\begin{equation}\label{winpix}\begin{array}{c}
w(\psi) = e^{\dot{\imath}\psi}\frac{\sin\psi}{\psi} \\
\overline{\lambda_{\ell}^m} = \frac{1}{\cos\theta_p -
\cos\theta_{p+1}} \int_{\cos\theta_p}^{\cos\theta_{p+1}}
\lambda_{\ell}^m(\cos\theta)d(\cos\theta) \, ,
\end{array}\end{equation}
From the last line in (\ref{almcomp}), the $b_m$ can be obtained by the
inverse sum:
\begin{equation}
b_m = \sum_q T_{pq}e^{-\dot{\imath}mq2\pi/N_p} \, ,
\end{equation}
which can be done using a Fast Fourier Transform.  The
sum (\ref{bmdef}) can also be inverted to obtain the $a_{\ell m}$'s:
\begin{equation}
a_{\ell m} = \frac{\sum_{p=1}^{n_{rows}} b_m w^{*}(\pi m/N_p)
\overline{\lambda_{\ell}^m}}{\sum_{p=1}^{n_{rows}} w^{*}(\pi m/N_p)
w(\pi m/N_p) N_p} \, .
\end{equation}
To obtain this last equation, we took the approximation that the pixelized
spherical harmonics $w(\pi m/N_p)\overline{\lambda_{\ell}^m}$ are
orthogonal, which is valid at low $\ell$.

\subsection{COBE Normalisation}\label{cobenorm}

To infer the cosmic string linear energy density, 
we use the COBE angular correlation function because, on the
scales of interests, it is good as that of WMAP~\cite{halpern2003}.
This also enables direct comparisons with previous work.

To normalize to COBE, we follow~\cite{allen96} and compute the angular
correlation function using
\begin{equation}\label{cthetaexp}
C(\theta, \theta_s) = \frac{1}{4\pi}\sum_{\ell} (2\ell
+1)|W_{\ell}|^2|B_{\ell}|^2C_{\ell}P_{\ell}(\cos\theta) \, ,
\end{equation}
where we use the following smoothing function:
\begin{equation}
W_{\ell}(\theta) = \exp\left(-\frac{\ell(\ell +1)}{\log
2}\sin^2(\theta/4)\right) \, ,
\end{equation}
where $\theta$ is the smoothing scales.  We also use the COBE beam
profile $B_{\ell}$ given in~\cite{wright94}.We then compare the result
with the value obtained by the four year COBE maps~\cite{banday97},
$C(0^{\circ},10^{\circ}) = (29\pm 1\mu K/T)^2$ to infer the value of
$G\mu/c^2$.  In the previous equation, $T$ is the mean temperature of
the CMB also obtained from COBE~\cite{bennett96} $T=2.726\pm 0.002K$.

\section{Results}

\subsection{Cosmic String Simulations}\label{sec_sim}

We created a time series of cosmic string realizations using 
the Allen-Shellard network simulation \cite{allen90} in a variety of
flat FRW
universes with $\Omega_m + \Omega_{\Lambda} =1$, where $\Omega_m$
includes the contribution of dark matter and baryons.  These string
networks were then used as 
sources for the CMB map-making pipeline.  For all the simulations, 
we chose $h =
0.72$ and $\Omega_bh^2 = 0.02$.

Two string simulations for
$\Omega_{\Lambda} = 0$ and $\Omega_{\Lambda} = 0.7$ were computed with
very high resolution with over 3 million points (16 ponts per
correlation length) and spanning a dynamic range of 5 in conformal
time, probing the Universe back to $z=25$ and $z=30$ respectively.
All string loops were retained in the simulation to ensure overall
network energy-momentum conservation, which was preserved at the level
of 1\%.  Point-joining was used only to ensure a consistent physical
resolution, i.e. removing very short segments formed through
reconnection.  The simulations therefore retained their full
small-scale structure.  These were the costliest part of the pipeline
and took many days to perform on the COSMOS supercomputer
(approximately 2000 hours of CPU time).  This improves upon previous
work with string networks in an expanding Universe, which was
performed with fewer points per correlation length, fixed horizon
resolution and a separate treatment of loops~\cite{allen96,allen97}.

In addition, we performed a series of smaller simulations (1 million
points) for different values of the cosmological constants,
specifically, $\Omega_{\Lambda} = 0,\, 0.2,\, 0.4,\, 0.55,\, 0.7,\, 0.85$.
These simulations had identical initial conditions, so comparisons
could minimize the effects of cosmic variance.

The map-making pipeline was tested at resolutions of $128^3$ and
$256^3$.  The huge data storage requirements with string
networks, SVT-decomposed energy-momentum tensor grids, Boltzmann code
output plus checkpointing files meant disk space approaching 1Tbyte
was required for $256^3$. Further parallelisation issues
remain to be resolved for large memory runs.

\begin{figure}
\resizebox{\columnwidth}{!}{\includegraphics{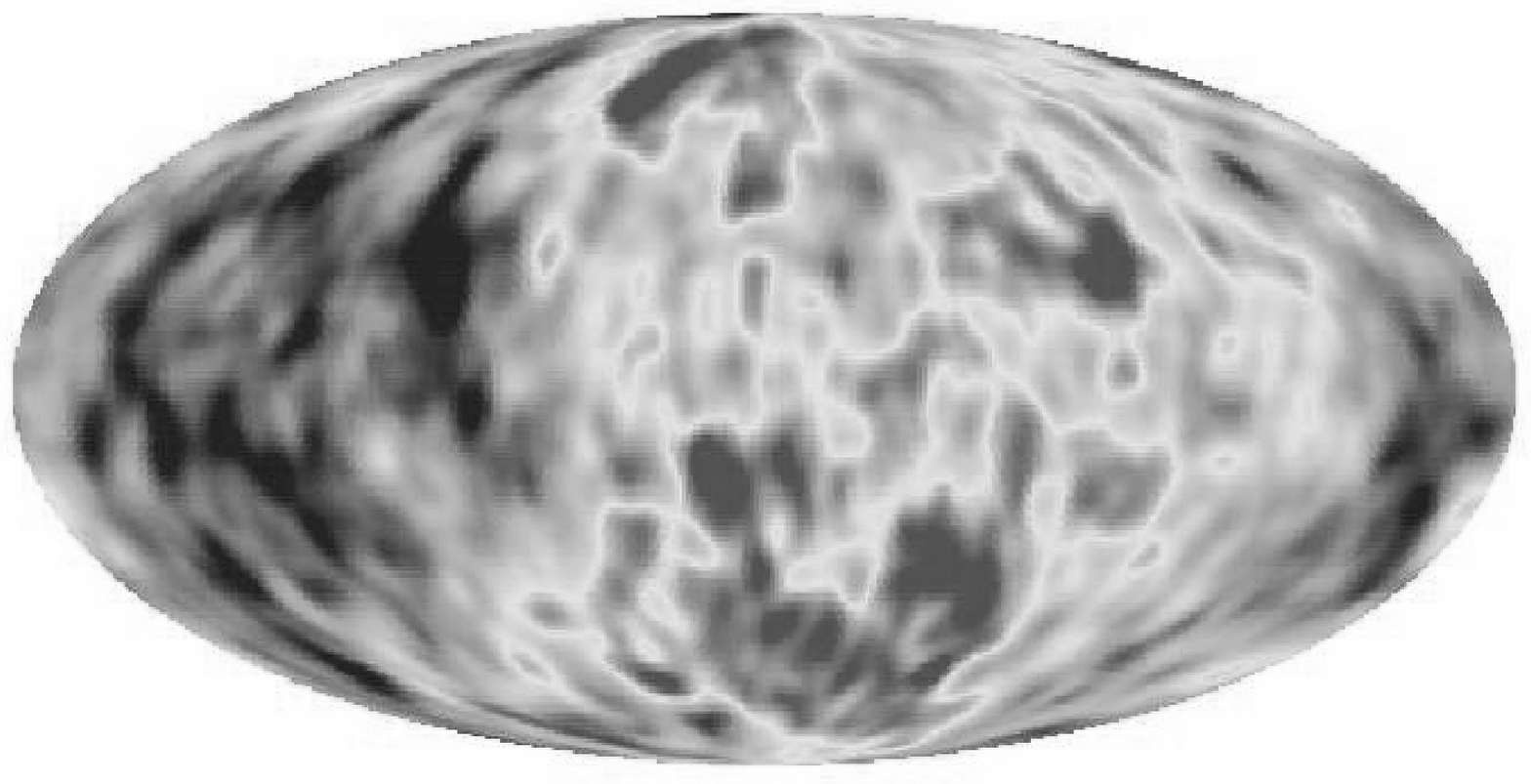}}
\resizebox{\columnwidth}{!}{\includegraphics{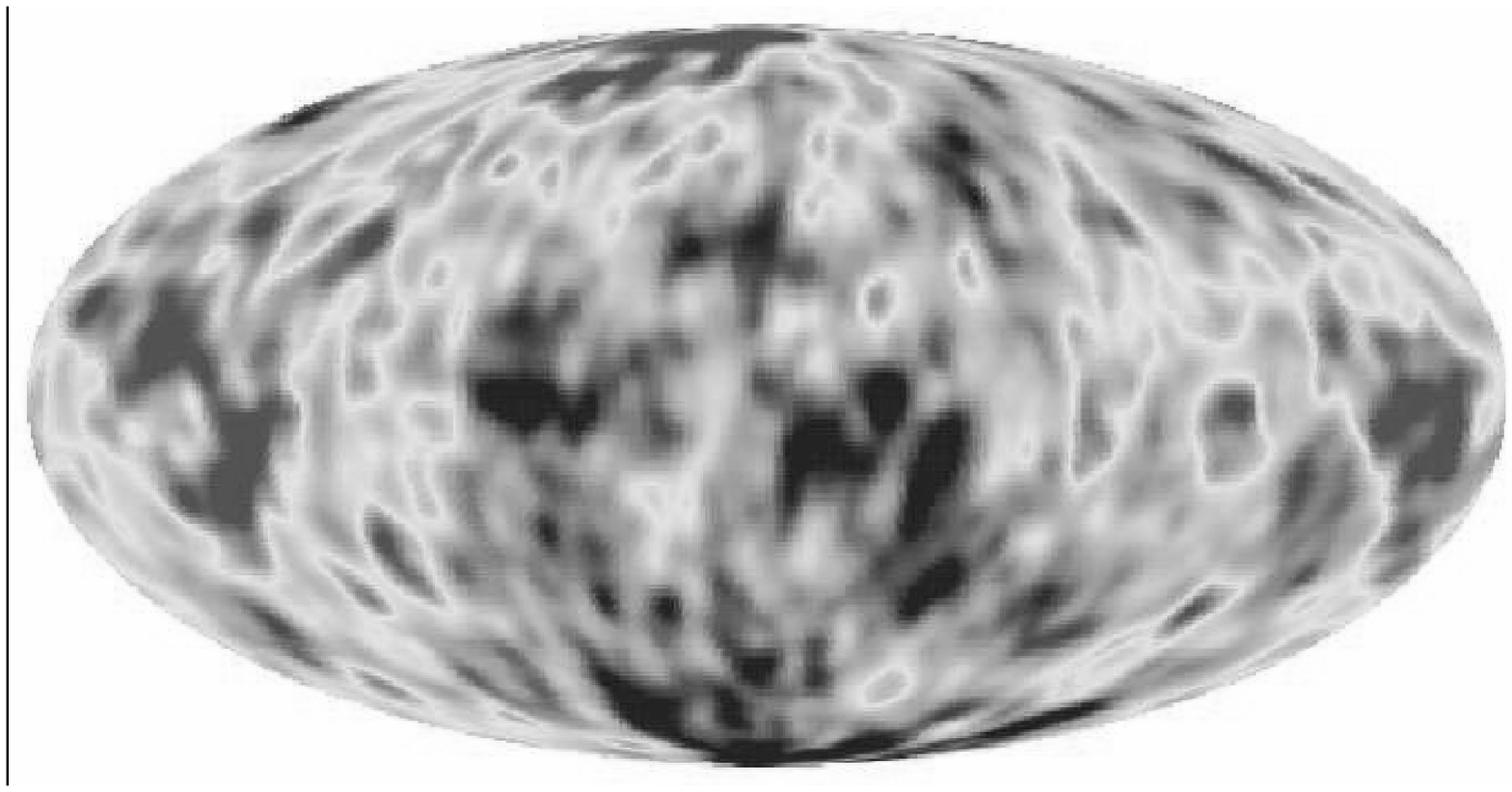}}
\caption{All-sky string-induced maps for a flat FRW Universe with 
$\Omega_{\Lambda} = 0$ and. $0.7$}\label{maps}
\end{figure}

\begin{figure}
\resizebox{\columnwidth}{5cm}{\includegraphics{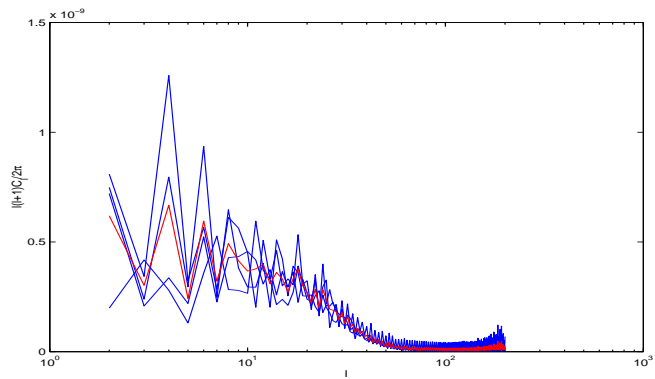}}
\caption{Individual power spectra for the 
$\Omega_{\Lambda} = 0$ run in blue with the average in
red.}\label{cl_all_fig}
\end{figure}

\begin{figure}
\resizebox{\columnwidth}{!}{\includegraphics{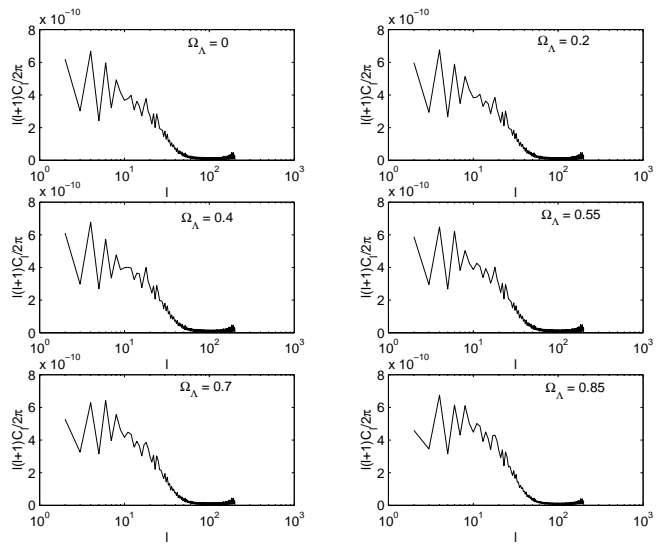}}
\caption{Average power spectra for each run.}\label{cls_all_fig}
\end{figure}

In figure~\ref{maps}, we present 
realizations for the $\Omega_{\Lambda} = 0$ and $0.7$ runs.
The angular power spectra for each
of these runs are presented in figure~\ref{cls_all_fig}.  We also show
four spectra obtained from the $\Omega_{\Lambda} = 0$ run with their
average in figure~\ref{cl_all_fig}.  All these spectra are normalised
to COBE.

\subsection{Normalization to COBE}

The power spectra illustrated in figures~\ref{cl_all_fig}
and~\ref{cls_all_fig} show a roughly scale invariant plateau for
$\ell\lsim 20$ with a gradual fall-off to an insignificant signal
beyond $\ell\gsim 50$.  This fall-off is a consequence of the limited
dynamic range of the string simulations and not of the resolution at
which the Boltzmann evolution was performed (though there is some
influence from the interpolation and smoothing of the string network
onto the grids.

Previous CMB work with string networks over a much larger
dynamic range has demonstrated that the main anisotropies relevant to COBE 
are
generated at redshifts $z\lsim 20$~\cite{allen96}, which is within
the range of the present work.  The reason for this can be seen from
studies of the unequal-time correlators
(UETC's)~\cite{albrecht96,wu2002}, which  peak on scales well below the horizon
 $(\lambda_{max}< d_H/3$).  We are confident therefore that our
simulations include the primary contributions on COBE scales.

This scale invariance on large angular scales is consistent with the
COBE data and allows us to normalize our results accordingly, as
discussed above, providing a constraint on the
energy density of strings.
Our normalization for the flat CDM model,
\begin{equation}
\frac{G\mu}{c^2} = (0.7\pm 0.2)\times 10^{-6} \, ,
\end{equation}
has an error reflecting the variance of our results, not systematic
effects which may be comparable.  This result is lower but consistent
with the previous result~\cite{allen96}:
\begin{equation}
\frac{G\mu}{c^2} = (1.05^{+0.35}_{-0.2})\times 10^{-6} \, .
\end{equation}
Our result is significantly lower than the one obtained
in~\cite{allen97}, $G\mu/c^2 = 1.7\times 10^{-6}$ from a single
simulation.  This work however focused on small angle anisotropies and
did not study sample variance at large angles.  Other flat space
approximations and semi-analytic estimates for local string networks
generally have found a higher normalization than ours,
e.g. $G\mu/c^2 = (1.5\pm 0.5)\times 10^{-6}$~\cite{bennett92},
$G\mu/c^2 = (1.7\pm 0.7)\times 10^{-6}$~\cite{perivolaropoulos93} and
$G\mu/c^2 = 2$~\cite{coulson94}.

However, there is a key reason why the present work is a significant
advance over these previous analyses.  Apart from now including all
the relevant physics, this analysis uses the highest resolution string
simulations to date.  The initial points per correlation length have
been chosen at levels known to preserve small scale structure
accurately~\cite{martins2001} and the simulations are evolved at fixed
physical resolution.  Although the simulations used in~\cite{allen96}
spanned a larger dynamic range, $z<100$ (compared to our $z<25$), it
did so by maintaining fixed horizon resolution through point-joining
and smoothing.  Subsequent work has shown that at least comoving
resolution is required if we are to hope for a satisfactory treatment
of string wiggliness.  This implies that the previous work did not
adequately account for the renormalized string energy per unit length,
which in the matter era is $\tilde{\mu} \simeq 1.4\mu$.  Such a factor
would tend to increase the string anisotropy, thus lowering the COBE
normalization of $G\mu/c^2$ (though in a non-trivial manner).  

\subsection{Effect of the Cosmological Constant}

\begin{figure}
\resizebox{\columnwidth}{!}{\includegraphics{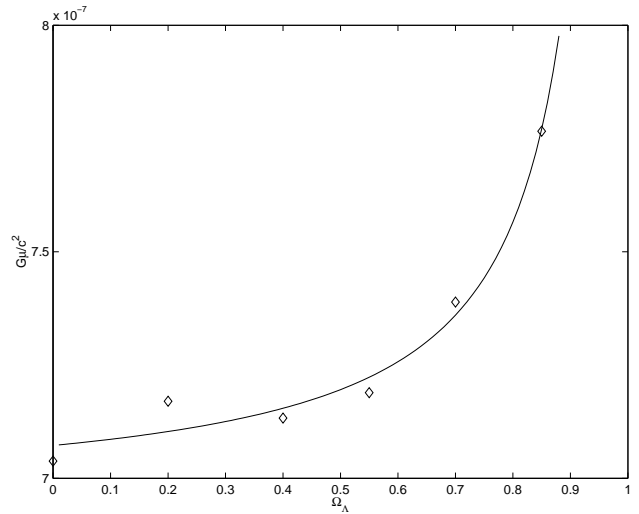}}
\caption{String linear energy density obtained from the COBE
normalisation.  The solid line is the curve (\ref{gmu_eq}).}\label{gmu_fig}
\end{figure}

The influence of $\Omega_{\Lambda}$ on the COBE normalization,
illustrated in figure~\ref{gmu_fig} is relatively small: for the
popular value of $\Omega_{\Lambda} = 0.7$ we obtained $G\mu/c^2 =
(0.74\pm 0.20)\times 10^{-6}$, only 6\% higher than for the flat CDM
model.

\begin{figure}
\hspace{1.5cm}
\resizebox{\columnwidth}{3.5cm}{\includegraphics{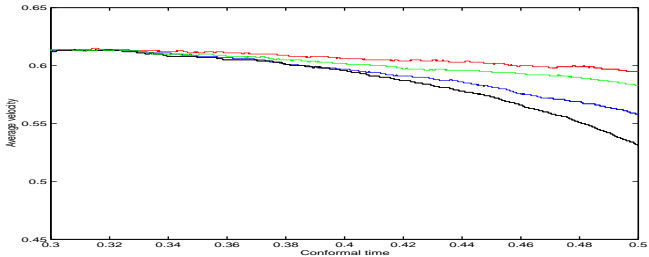}}
\caption{Average rms long string velocity for the last half of the
string network simulations for $\Omega_\Lambda = 0$ (red),
$\Omega_\Lambda = 0.4$ (green), $\Omega_\Lambda = 0.7$ (blue) and
$\Omega_\Lambda = 0.85$ (black).  Even the extremal $\Lambda$-model
shows only a relatively small 11\% decline in the average string
velocity by the present day.  (Relative conformal time is plotted with
$\eta_0 = 0.5$.)}\label{stringvel_fig}
\end{figure}

The reason for the reduced anisotropy in $\Lambda$-models is fairly
clear.  As the Universe becomes vacuum dominated at late times, the
expansion rate increases, affecting the Hubble damping term
$\dot{a}/a$ in the string equations of motion, thus lowering their
average velocity as can be seen from figure~\ref{stringvel_fig}.  
As string velocities are the primary cause of CMB anisotropies, there
will be a net reduction in $\Delta T/T$.

The smallness of this effect is somewhat surprising, but is explained
by the late redshift of vacuum domination.  In the $\Omega_{\Lambda} =
0.7$, this occurs at $z_{\Lambda}\simeq 0.33$, which, in the context of
our simulations, implies that $\Lambda$ has a significant effect on only
the later stages of the simulation.  Comparing
the $\Omega_{\Lambda} = 0.85$ model with the flat CDM model in
figure~\ref{cls_all_fig}, there appears to be a slight relative fall-off in
the average power towards $\ell=2$, which is consistent with this
picture.  However, this effect is likely to be swamped by cosmic
variance.

We have found that our data points are well fitted by
\begin{equation}\label{gmu_eq}
\frac{G\mu}{c^2} = \left(0.695 + \frac{0.012}{1-\Omega_{\Lambda}}
\right)\times 10^{-6} \, .
\end{equation}
This analytic fit is motivated by the asymptotic limit in which string
velocities should vanish as the network is frozen and conformally
streched in an extreme $\Lambda$-model: $\Omega_\Lambda \rightarrow 1$
and $\Omega_m\rightarrow 0$.
Our results are summarized in table~\ref{gmu_tab} and plotted in
figure~\ref{gmu_fig} along with the fit (\ref{gmu_eq}).

\begin{table}
\begin{center}
\begin{tabular}{cc}
$\Omega_{\Lambda}$ & $G\mu/c^2\times 10^{-6}$  \\
 0.00  &   0.7038  $\pm$  0.1947 \\
 0.20  &   0.7170  $\pm$  0.1926 \\
 0.40  &   0.7133  $\pm$  0.1989 \\
 0.55  &   0.7189  $\pm$  0.1902 \\
 0.70  &   0.7389  $\pm$  0.1990 \\
 0.85  &   0.7766  $\pm$  0.1879
\end{tabular}
\end{center}
\caption{String linear energy density obtained from the COBE
normalisation.}\label{gmu_tab}
\end{table}

\section{Summary}

In this paper, we presented large angle maps of CMB fluctuations
seeded by networks of cosmic strings in flat FRW universes with a cosmological
constant. From the COBE data, we have obtained a constraint on the 
string linear energy 
density $G\mu/c^2 \lsim 0.7\times 10^{-6}$ which is lower 
than previous work because 
string small-scale structure is incorporated more accurately in the 
network simulations.  
We were able to find the dependence of the 
string density $\mu$ as a function of the cosmological constant, 
obtaining a good fit with a simple
semi-analytic formula.    Given the uncertainties in the overall
normalization, these results should provide an adequate means by which
to characterise the effects of $\Omega_{\Lambda}$ on cosmic string
models.

Although current CMB data have shown that cosmic strings cannot be the
dominating source of CMB anisotropies (e.g.~\cite{wmap1}), they can
nonetheless be present albeit at a lower energy scale.  For example,
they are copiously produced at the end of brane
inflation~\cite{sarangi2002, pogosian2003}.  With this in mind, our
normalisation appears low when compared with the value inferred from
the possible detection of a gravitational lensing event by a cosmic
string: $G\mu /c^2 \approx 0.4\times 10^{-6}$~\cite{sazhin2003}.  This
linear mass density would be somewhat higher than allowed in the
standard cosmic string scenario considered in this paper.

\section*{Acknowledgements}

We are grateful for useful discussions with Gareth Amery, Richard Battye, 
Martin Bucher, Carlos Martins, Proty Wu and Rob Crittenden.
The Allen-Shellard string simulation was used to generate the networks
used as sources in this paper \cite{allen90}.
This work was supported by PPARC
grant no.  PPA/G/O/1999/00603.  All simulations were performed on COSMOS,
the Origin 3800 supercomputer, funded by SGI, HEFCE and PPARC.

\bibliography{refs,myrefs,addref,wmap}

\end{document}